# Charge inversion and topological phase transition at a twist angle induced van Hove singularity of bilayer graphene

Youngwook Kim,[†] Patrick Herlinger,[†] Pilkyung Moon,[‡,§] Mikito Koshino,[∥] Takashi Taniguchi,[⊥] Kenji Watanabe,[⊥] and Jurgen. H. Smet[*,†]

[†]*Max-Planck-Institut für Festköperforschung, 70569 Stuttgart, Germany.*

[‡]*New York University, Shanghai 200120, China*

[§]*NYU-ECNU Institute of physics at NYU Shanghai, Shanghai 200062, China*

[∥]*Department of Physics, Tohoku University, Sendai 980-8578, Japan*

[⊥]*National Institute for Materials Science, 1-1 Namiki, Tsukuba, 305-0044, Japan*

**Abstract**

van Hove singularities (VHS's) in the density of states play an outstanding and diverse role for the electronic and thermodynamic properties of crystalline solids. At the critical point the Fermi surface connectivity changes, and topological properties undergo a transition. Opportunities to systematically pass a VHS at the turn of a voltage knob and study its diverse impact are however rare. With the advent of van der Waals heterostructures, control over the atomic registry of neighboring graphene layers offers an unprecedented tool to generate a low energy VHS easily accessible with conventional gating. Here we have addressed magnetotransport when the chemical potential crosses the twist angle induced VHS in twisted bilayer graphene. A topological phase transition is experimentally disclosed in the abrupt conversion of electrons to holes or vice versa, a loss of a nonzero Berry phase and distinct sequences of integer quantum Hall states above and below the singularity.

VHS's have been frequently invoked as the source of rich physics.[1] We exemplary refer here to their influence on superconductivity. Near the van Hove filling the electron−phonon coupling may get enhanced thereby enabling superconductivity or affecting its stability.[2] They can also mediate nematic or other deformations of the Fermi surface in the presence of interactions and trigger Landau type symmetry breaking phase transitions through a Pomeranchuk instability.[3] In bulk materials the chemical potential is not readily tunable across a wide enough energy range to explore the physics by a VHS in a systematic manner. Conventional and electrolyte gating techniques applicable to low-dimensional materials seem suitable for such purposes. Graphene has been considered a promising candidate material for such studies and, stimulated by the controversially debated prospect of turning it into a superconductor, has been looked at intensively up to high densities.[4,5] Its $\pi$-conduction and valence band possess a VHS due to the merger of Dirac cone states emanating from the **K** and **K´** symmetry points in the Brillouin zone. However, its energy is in essence determined by the large intersublattice hopping parameter of 2.7 eV apparently placing it out of reach for even the most potent in situ controllable gating techniques based on electrolytes.

From high-temperature superconductors it is well-known that VHS's are common in quasi two-dimensional layered materials where they arise as a result of the interlayer coupling.[6−8] In van der Waals heterostructures this interlayer coupling can be tuned by changing the atomic registry between neighboring layers.[9−11] It has been confirmed in scanning tunneling microscopy and spectroscopy that in twisted bilayer graphene where the crystallographic axes of the neighboring layers are misaligned by a rotation angle $\theta$, a VHS exists across a broad range of twist angles.[12−14] These occur at controllable and, most of all, lower energies than in single layer graphene. This makes the singularity accessible with conventional field effect gating methods.

Figure 1 illustrates the band structure and density of states of a twisted bilayer of graphene with a 2° misalignment of the top layer. The calculation procedure has been outlined in the Method section. As a result of the misalignment the Dirac cones of the top layer get displaced in *k*-space by an amount equal to $|\mathbf{G}|/\sqrt{3} \times 2\sin(\theta/2) \cong |\mathbf{G}|/\sqrt{3} \times 0.035$, where $\theta$ is the twist angle of 2° and $|\mathbf{G}|$ is the size of the reciprocal vector of monolayer graphene. At low energies, states of the displaced cones of the bottom and top layer at $\mathbf{K}_t$ and $\mathbf{K}_b$ are only tunnel coupled, however eventually a hybridization is inevitable generating a VHS at much lower energy since the distance between the displaced cones is only a fraction of the distance between the $\mathbf{K}$ and $\mathbf{K}'$ symmetry points of the Brillouin zone of a single layer. Due to the Moiré superstructure the first Brillouin zone shrinks by a factor $(2\sin(\theta/2))^2 \sim 0.0012$ (=0.12 %) as illustrated in Figure 1a and the density to fill the lowest bands is reduced accordingly. This lowest conduction and valence band is compressed down to 170 meV as compared with 6 eV for single layer graphene and the van Hove energy equals 24.5 meV only. A three-dimensional color rendition visualizing the Dirac cones and the saddle point generating the VHS is displayed in Figure. 1b. The symmetry points in the reduced Brillouin zone are referred to as the $\overline{\mathbf{K}}$ and $\overline{\mathbf{K}}'$ symmetry points in analogy to the original Brillouin zone. We note however that for single layer graphene the two Dirac cones at the $\mathbf{K}$ and $\mathbf{K}'$ symmetry points are related by time reversal symmetry and the Berry curvatures are naturally opposite. Here, the Dirac cones at $\overline{\mathbf{K}}$ and $\overline{\mathbf{K}}'$ are not time related, and the symmetry of the interlayer hopping terms ensures that the Berry curvatures are identical instead. There is another lowest energy conduction band which emerges from the $\mathbf{K}'_t$ and $\mathbf{K}'_b$ Dirac cones of the single layers. It is mirror symmetric to the band from the $\mathbf{K}_t$ and $\mathbf{K}_b$ Dirac cones as seen in the iso-energetic contour plots in Figure 2. States of both conduction bands do not mix. They are hereafter referred to as the $\mathbf{K}$ and $\mathbf{K}'$ conduction bands. On the contour plots the winding number for each class of orbits has been marked. The contour at the van Hove energy is plotted in red. When crossing this energy the topology of the Fermi surface is altered. The winding number drops for -1 or +1 (depending on the conduction band) to 0 since higher energy contours encircle the $\bar{\Gamma}$ symmetry point where no Berry curvature exists. Arrows along the contour lines indicate whether orbits are executed clock- or anticlockwise for a chosen perpendicular field orientation. The remainder of this manuscript addresses the key experimental signatures of the gate voltage induced topological changes when crossing the van Hove energy.

For this study a van der Waals heterostructure has been manufactured using the stamp pick-up transfer method.[15] The stack was placed on top of a silicon substrate covered by a thermally grown oxide. It was composed of the following sequence of layers: a graphitic back-gate, h-BN, a twisted bilayer of graphene, h-BN, a graphitic top gate and finally h-BN. A schematic of this layer sequence together with optical microscope images of the actual device and contact configuration are shown in the Supporting Information (Figure S1). The graphitic gate layer and h-BN layer at the bottom have a total thickness well exceeding 10 nm thereby providing a support of exceptional flatness which has been demonstrated to reduce strain fluctuations, the main source of scattering and also the culprit for density inhomogeneities.[16] The bottom graphite layer also effectively screens any disorder potential originating from polar or charged impurities on top or in the thermal oxide, which represents an additional measure to enhance the sample quality.[17,18] The gate voltage dependence of the four terminal longitudinal resistance $R_{xx}$ is shown at the bottom of Figure 3a. The charge neutrality point (CNP) associated with the $\bar{K}$ and $\bar{K}'$-Dirac cones in both bands centered around zero energy (see Figure 1) is located at a gate voltage of 0.015 V. The resistance peak exhibits a full width-half maximum of 0.015 V corresponding to a residual charge carrier density from electron-hole puddles[19] of $5\times10^{10}$ cm$^{-2}$. This represents to the best of our knowledge a record value for a twisted bilayer[20-23] and competes with state-of-the-art Bernal stacked bilayer graphene samples.[17,24-26] The top- and back-gate were unintentionally short circuited. Because the h-BN gate dielectric layers do not have the exact same layer thickness (approximately 6 and 9 nm for the hBN layer above and below the twisted bilayer, respectively), the application of a gate voltage may induce a small transverse electric field across the twisted graphene layers (for further information and an estimate of the size of the transverse electric field we refer to section S1 in the Supporting Information). The two monolayer graphene flakes were selected to possess edges that reflect the hexagonal crystal structure in order to be able to align them at a suitable angle during the stacking procedure. Here an angle of 2° to 3° was aimed for. To avoid additional Moiré superstructures from van der Waals interactions between the graphene monolayers and the top or bottom h-BN, a large twist angle of approximately 30° between each of the graphene monolayers and the nearest h-BN layer was chosen. More details of the sample fabrication procedure are deferred to the Method section.

Apart from the sharp peak in the longitudinal resistance due to the $(\overline{\mathbf{K}}, \overline{\mathbf{K}}')$-CNP, two additional broader satellite features appear near gate voltages of ± 1.9 V in Fig. 3**a**. Their temperature dependence is distinct from the main charge neutrality peak, which vanishes more rapidly with temperature (see also Figure S2 for additional temperature dependent data). We assert that this renewed rise and subsequent drop in resistance near ± 1.9 V is a manifestation of the complete filling of the bands centered around zero energy and the subsequent filling of the next set of mini-bands formed at the $\overline{\Gamma}$-point during zone folding (see Figure 1 and Figure S3) as a result of the expansion of the real space unit cell by the Moiré superlattice potential. The location of these resistance features will hereafter be referred to as the secondary or $\overline{\Gamma}$ CNPs. A closer inspection reveals that these resistance features are composed of two peaks (marked with two different symbols in Figure S2), both of which display a significant temperature dependence above 100 K only. This temperature dependence is tentatively attributed to the necessity of phonon assisted tunneling for indirect transitions from the $\overline{\mathbf{\Gamma}}$-point of the lowest band to the absolute energy minimum of the second band along the $\overline{\mathbf{\Gamma}}$ - $\overline{\mathbf{M}}$ line (see Figure 1 and 2 as well as Supporting Information S2 for a more detailed description). A second contribution may stem from direct transitions near the $\overline{\mathbf{\Gamma}}$-point, which require overcoming an energy gap of approximately 10 meV.

The complete filling of the lowest bands is also confirmed in magnetotransport data. Color renderings of both the longitudinal as well as the Hall resistance as a function of the gate voltage and the applied perpendicular magnetic field $B$ are shown in panel a and b of Figure 3, together with some single line traces. While the Landau fan emanating from the $(\overline{\mathbf{K}}, \overline{\mathbf{K}}')$-CNP dominates these figures, secondary Landau fans with their origin near ± 1.9 V develop as the magnetic field is raised. The supporting section includes a two-dimensional plot of the derivative $dR_{xx}/dV_g$ (Figure S4, section S3) to bring out the Landau fans even more clearly. The Hall resistance switches from electron to hole conduction or vice versa near ± 0.9 V. Since these secondary CNPs indicate complete filling of the lowest energy bands, the required gate voltage to reach them enables us to estimate the size of the superlattice unit cell and the corresponding twist angle (for details see section S4 in the Supporting Information). They are approximately equal to 7 nm and 2º, respectively. This justifies a posteriori the choice of the twist angle for the band structure calculations highlighted in Figure 1 and 2. Somewhat related behavior in the resistivity has been reported for single layer graphene deposited on top of h-BN with a twist angle less than 1° where secondary Dirac cones emerge in the band structure;[9-11] however, there are significant disparities. For instance, in monolayer graphene/h-BN stacks the strength of the resistance peaks associated with the valence and conduction band secondary Dirac cones are very different. In our case these satellite resistivity peaks at the $\overline{\mathbf{\Gamma}}$-CNPs exhibit near perfect electron-hole symmetry. Unlike for the graphene/hBN system, the Hamiltonian of twisted bilayer graphene possesses symmetry properties[27] which prevent the development of a strong asymmetry between the electron and hole energy spectra in the framework of the effective mass approximation. This absence of strong asymmetry is apparent from the band structure calculations displayed in Figure 1 (see also Figure S3). The available differences mainly originate from the inclusion of the next-to-nearest neighbor hopping terms in the tight binding model.

Very striking is the conversion of the charge carriers from hole to electron-like or vice versa observed in panel **a** and **b** of Figure 3 near ± 0.9 V, almost exactly midway between the primary CNP at zero gate voltage and the secondary CNPs (± 1.9 V). This apparent rotation reversal takes place within the band. We assert that it is one manifestation of the continuum topological Lifshitz transition due to a change of the Fermi surface topology when the chemical potential crosses the twist angle induced VHS[28]. We refer to Figure 2 illustrating the iso-energetic contours below and above the VHS which indeed predicts the reversal from anti-clockwise to clockwise rotation.

The winding number, whose sign also reveals the helicity, has been included for each category of closed orbits. Below the VHS both conduction bands possess two iso-energetic contours with identical winding numbers: 1 or -1. The sign is opposite for the equivalent orbits of different conduction bands reflecting the original time reversal symmetry requirement between the **K** and **K´** Dirac cone states of the single graphene. Above the Fermi surface singularity, only one iso-energetic contour per conduction band remains, and the winding number drops to zero in view of the absence of Berry curvature around the $\bar{\Gamma}$ point. We conclude based on these properties of the iso-energetic contours of the lowest conduction bands that a transition across the van Hove Fermi surface singularity should also be accompanied by the following experimental observations:

(i) At low energies close to the $(\overline{\mathbf{K}}, \overline{\mathbf{K}}')$-CNP the single layer Dirac cone states barely interact as they are only tunnel coupled. The iso-energetic contours enclose a single Dirac point and the Berry curvature is identical to that in monolayer graphene. Accordingly, the sequence of incompressible integer quantum Hall (QH) states should initially at low densities correspond to the monolayer graphene filling factor sequence multiplied by two to account for the existence of two layers and hence two conduction bands: $\nu = \ldots -20, -12, -4, 4, 12, 20, \ldots$. The hybridization of their states at higher energies is responsible for the appearance of the VHS, and the degeneracy of the iso-energetic contours drops from 2 to 1 for each conduction band. In the vicinity of the $\overline{\Gamma}$ secondary CNPs when the chemical potential exceeds the VHS, we anticipate a sequence of integer QH states identical to that of Bernal stacked bilayer graphene instead: $\nu = \ldots, -12, -8, -4, 4, 8, 12, \ldots$ The winding number discrepancy of 0 for orbits encompassing the $\overline{\Gamma}$-point in twisted bilayer graphene and 2 for conduction band states of Bernal stacked bilayer graphene encircling the $\mathbf{K}$ or $\mathbf{K}'$ symmetry points is indistinguishable and therefore without observable consequences. These sequences assume that neither spin nor layer degeneracies are broken by symmetry breaking fields such as the externally applied magnetic field (Zeeman splitting) and the transverse electric field (layer splitting).

The condensation into an integer QH state appears in the magnetotransport data of Figure 3a as a dark line with constant Landau level filling slope extrapolating to the zero field resistivity peaks associated with the primary and the secondary CNPs at $(\overline{\mathbf{K}}, \overline{\mathbf{K}}')$ and $\overline{\Gamma}$. The fillings have been marked at the top and indeed confirm these sequences. In panels c and d line traces of the Hall resistance allow verifying the values of the Hall plateaus as additional, redundant evidence. The QH effect emanating from the primary CNPs starts at 0.3 T, and broken symmetry states in the zeroth Landau Level can be discerned from 3 T. The layer degeneracy gets lifted due to the gating induced transverse electric field above gate voltage

±0.4 V. From here on fainter lines corresponding to filling 8, 16 and 24 appear in addition (see also the plot of $dR_{xx}/dV_g$ in Figure S4).[28] The size of the transverse electric field can be estimated from the thickness difference of the hBN dielectric layers separating the twisted bilayer from the graphitic gate layers as discussed in section S1 of the Supporting Information.

(ii) The drop of the winding number from ±1 to 0 implies that the Berry phase extracted from quantum oscillation extrema should change from ±$\pi$ to 0. Figure 4a and b displays fan diagrams for Shubnikov-de Haas oscillations near the primary (panel b and d) and secondary CNPs (panel a and c) obtained by plotting the inverse of the magnetic field value at which extrema occur as a function of the Landau index $n_L$. The slope of a linear fit to the data points gives the charge carrier density and the intercept of the linear fit with the level index axis multiplied by $2\pi$ yields Berry's phase. The phase extracted in this manner is plotted in panel e of Figure 4 for different gate voltages. It imposingly corroborates the theoretical expectation of a Berry phase change from ±$\pi$ to 0 due to the topological Lifshitz transition at the VHS. From the Shubnikov-de Haas oscillation period the charge carrier density can also be extracted. The results are summarized in section S5.

At the van Hove energy the cyclotron mass of the charge carriers is expected to diverge. In general the cyclotron mass can be extracted from a temperature dependent study of the Shubnikov-de Haas oscillations with the help of the Lifshitz-Kosevich formula.[29] Details of this procedure are outlined in the Supporting Information S5. The gate voltage dependence of the cyclotron mass obtained in this manner is contained in Figure 5. Included in this graph are the theoretically expected mass values from the band structure calculations in Figure 1. For the sake of comparison we also show the density independent mass expected in Bernal stacked bilayer graphene (BBG) as well as the square root dependent mass for monolayer graphene (MG). Experiment indeed reveals a substantial mass enhancement near the VHS. It reaches values five times larger than in Bernal stacked bilayer graphene, but the dependence is significantly broadened. The discrepancy between the theoretically calculated mass and the experimentally extracted mass should not come as a surprise. The theory assumes an ideal system without any disorder, scattering or a spatially varying electric field across the layers. Hence, any broadening, inevitable in experiment, is absent and the van Hove singularity produces a sharp unrealistic spike in the mass. The experimental mass is extracted from transport which averages across the entire sample. For instance, the existence of electron-hole puddles implies a spatial dependence of the chemical potential. From the field effect trace in Figure 1 we concluded that the electron and hole puddles near charge neutrality host on average a charge carrier density of $5\times10^{10}$ cm$^{-2}$. This disorder will generate average spatial variations of the chemical potential of the order of 5 meV which are expected to smear out significantly the mass singularity. We have added to Figure 5 a curve of the effective mass obtained after a convolution of the density of states with a Gaussian whose FWHM was varied. A reasonable energy broadening of 10 meV yields comparable mass values to those obtained in experiment.

We note that not only spatial variations of the chemical potential contribute to this inhomogeneous broadening. After all the electron-hole puddle landscape should originate from or be accompanied by a spatial variation of the transverse electric field as well. It causes a spatially dependent asymmetric doping of the layers and a small splitting of the Dirac cones of the two separate layers. This is bound to be a further cause of inhomogeneous broadening. Of course also the homogeneous transverse electric field due to the differences in the thickness of the dielectric layers creates an asymmetric doping and splitting of the cones. Manifestations of homogeneous asymmetric doping have been considered in detail in Ref. 30 and 31. Its influence should be particularly important near charge neutrality and it is plausible to assume that its incorporation would alleviate the discrepancy between theory and experiment in this region further.

We conclude that twisted bilayer graphene, in which the two layers have a small rotational mismatch, represents a powerful paradigm to study topological aspects of the band structure. Interlayer hopping results in a van Hove singularity at low energy that can readily be populated with conventional gating techniques, in contrast with the van Hove singularity in single layer graphene that remains out of reach even with electrolyte gating methods. The magnetotransport data in this work offer ample evidence for a topological Lifshitz phase transition at the singularity fully compatible with band structure calculations. This includes a transition of the Berry phase extracted from quantum oscillations, a transition of the sequence of integer QH states and an apparent reversal of the orbital motion in the Hall effect. Secondary charge neutrality features in the resistance prove the existence of mini-gaps. Their temperature dependences suggest contributions from direct as well as indirect phonon assisted transitions between adjacent mini-bands. This work should stimulate further studies of thermodynamic properties that can be strongly affected by a van Hove singularity, here at the turn of a voltage knob.

**Methods**

**Device Fabrication.** The fabrication of the van der Waals heterostructure consisting of graphite gates and two twisted graphene monolayers sandwiched between thicker h-BN layers was performed with a polypropylene carbonate (PPC) stamp based pick-up method[15]. The stamp itself was manufactured by spin coating PPC onto a bare silicon substrate. After baking at 80°C for 5 minutes, the PPC film was mechanically peeled off and transferred on top of a small, about one millimeter thick PDMS stamp which itself was attached to a glass slide. The starting graphene monolayers and h-BN flakes were mechanically exfoliated onto silicon substrates covered by a 90 nm thick thermally grown oxide layer. Atomically flat flakes were selected using dark field imaging. Only monolayer flakes with boundaries revealing the hexagonal crystal structure were selected. Such boundaries were then exploited to align the monolayers. The graphene and h-BN layers were picked up with the stamp mounted on a micromanipulator. In order to increase the transfer yield as well as decrease the number of bubbles and wrinkles, the substrate supporting the atomically flat layers was fixed onto an aluminum holder which was pre-heated to 40°C while the stamp was slowly approached. When one edge of the stamp touched the substrate, the temperature of the aluminum support, and hence also the substrate, was slowly raised at a rate of approximately 0.1°C/min to 45-50°C, while maintaining the micromanipulator and stamp at fixed height. Due to the thermal expansion of the aluminum support structure, the stamp eventually gets in touch with the target flake to be picked up. Subsequently, the temperature was decreased to 35°C and the stamp lifted. The same procedure was repeated to assemble the complete heterostructure consisting of a total of 7 flakes: the top h-BN, a top graphitic gate layer, the middle h-BN, the two twisted graphene monolayers, the bottom h-BN and a graphite back/gate. The stamp with van der Waals stack was then placed on a Si substrate covered by a 300 nm thick oxide. This substrate was heated very slowly to 120°C in order to release the stack from the stamp and reduce bubbles and wrinkles. To improve their flatness and remove PPC polymer residues, the heterostructures were put in a furnace at 500°C with forming gas atmosphere for one hour prior to further processing. The h-BN and graphene were etched in a $CHF_3$ and $O_2$ containing plasma while the final device geometry was protected with a 500 nm PMMA etching mask.[32] Electrodes were patterned with e-beam lithography using an 800 nm thick PMMA layer that served both as an etch protection and lift-off layer. In order to enhance the contact quality, an etch retreatment in the $CHF_3$ and $O_2$ plasma was performed just before the metal evaporation.[32] The contacts were composed of 70 nm of gold deposited on top of a 10 nm Cr adhesion layer. Both were evaporated thermally. The graphite top- and back-gates were unintentionally short circuited. Because the h-BN dielectric layers do not have the exact same thickness, the development of a small transverse electric field across the twisted bilayer upon increasing the gate voltage cannot be avoided.

**Theoretical model for electronic band structures of twisted bilayer graphene.** For the numerical calculation of the electronic band structures shown in Figures. 1, 2 and 5 of the main text and Figure S3 of the Supporting Information, we modeled the twisted bilayer graphene by a single layer of graphene overlaid on top of the other with a twist angle $\theta$, where $\theta = 2°$ is obtained from experimental data. We used a lattice period of $a = 0.246$ nm for each graphene layer and an interlayer distance $d$ of 0.335 nm. The electronic structures are calculated by the effective continuum model based on the Dirac equation in which the effect of the periodic potential is fully taken into account.[27] The interlayer coupling strength $u_0 = 0.11$ eV in this effective model is obtained by the Slater-Koster parametrization of the hopping integral.[28]

**Supporting Information**

Supporting information is available free of charge on the ACS Publications website at DOI: 10.1021/acs.nano-lett.6b0196

Device images and hBN layer thickness determination, Temperature dependence of the longitudinal resistance $R_{xx}$ at $B = 0$ T, Evidence for charge localization physics near the $\bar{\Gamma}$–CNP, Determination of the superlattice unit cell and twist angle, Carrier density analysis by Shubnikov-de Haas oscillations, Cyclotron mass extraction from Shubnikov-de Haas oscillations

**Corresponding Author**

*E-mail: j.smet@fkf.mpg.de

**Author contributions**

Y. K and J.H.S conceived the project. Y.K and P.H carried out device fabrication and Y.K performed electrical measurements and Y.K and J.H.S interpreted the data. P.M and M.K provided theoretical support. T. T and K.W synthesized the h-BN bulk crystal. Y.K, P.H, P.M, M.K and J.H.S co-wrote the manuscript.

**Notes**

The authors declare no competing financial interest.

**Acknowledgments**

We thank K. von Klitzing and J.S Kim for fruitful discussions and J. Mürter, Y. Stuhlhofer, S. Göres and M. Hagel for assistance with sample preparation. We extend our gratitude to M.

Lee, P. Gallagher and D. Goldhaber-Gordon for fruitful input about the van der Waals heterostructure assembly. This work has been supported by the graphene flagship and the DFG Priority Program SPP 1459. P.M acknowledges the support of the NYU Shanghai (Start-up Funds), NYU-ECNU Institute of Physics, and the NSFC Research Fund for International Young Scientists 11550110177. K.W. and T.T. acknowledge support from the Elemental Strategy Initiative conducted by the MEXT, Japan and a Grant-in-Aid for Scientific Research on Innovative Areas "Science of Atomic Layers" from JSPS.

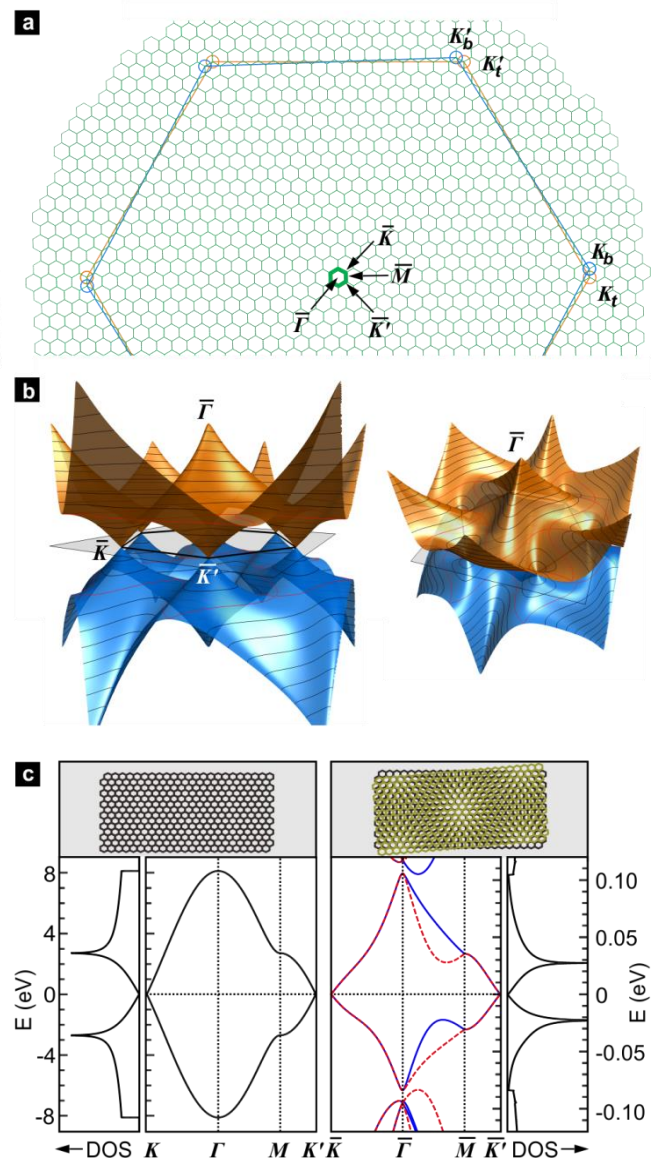

**Figure 1.** Brillouin zone, band structure and density of states of twisted bilayer graphene with an angle of 2°. (a) Brillouin zone. Thick green lines at the center indicate the reduced Brillouin zone. The symmetry points $\bar{\mathbf{K}}, \bar{\mathbf{K}}', \bar{\mathbf{\Gamma}}$ and $\bar{\mathbf{M}}$ have been marked. The original Dirac points of the bottom graphene monolayer ($\mathbf{K}_b$ and $\mathbf{K}_b'$) and the displaced Dirac points of the top layer ($\mathbf{K}_t$ and $\mathbf{K}_t'$) are also shown. (b) Side and top view of a 3D color rendering of the *K*-conduction band centered around zero energy. (c) Right: Two dimensional representation of the dispersion of the $\bar{\mathbf{K}}$ and $\bar{\mathbf{K}}'$-conduction bands (dashed red and solid

blue, respectively) and the corresponding density of states. Left: Density of states and dispersion of the lowest conduction band in monolayer graphene for comparison.

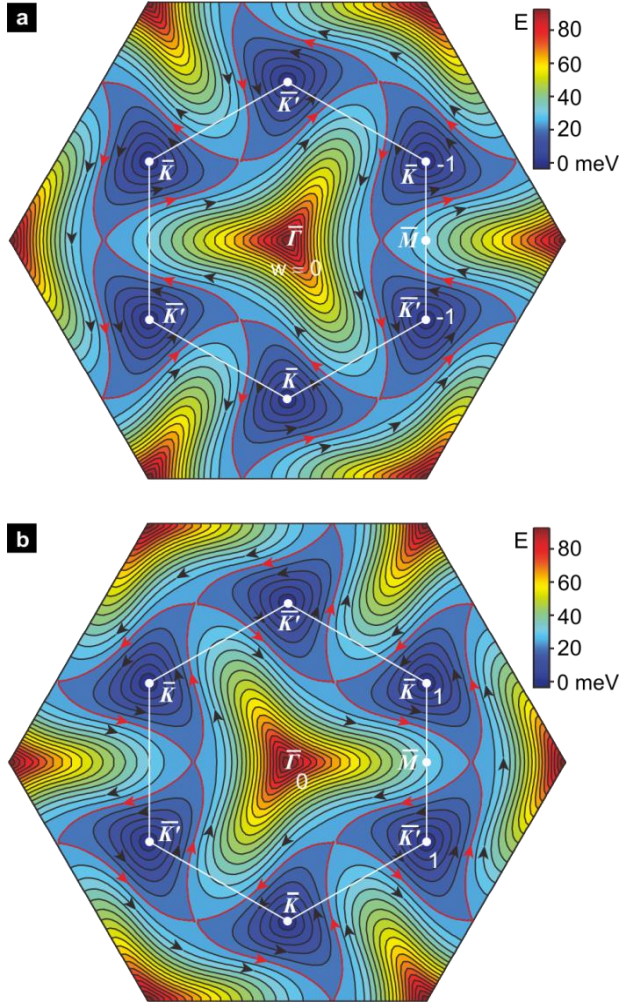

**Figure 2.** Contour plots of the lowest conduction bands of twisted bilayer graphene with θ = 2°. (a) Isoenergetic contours of the **K**-conduction band. The rotation direction and winding number of the orbits encompassing the $\bar{\Gamma}$, $\bar{K}$ and $\bar{K}'$ symmetry points have been indicated by arrows and numbers (0 or ±1), respectively. The van Hove singularity occurs at 24.54 meV. The contour at this energy is shown in red. Contours are 4 meV apart. (b) The same but for the **K'**-conduction band.

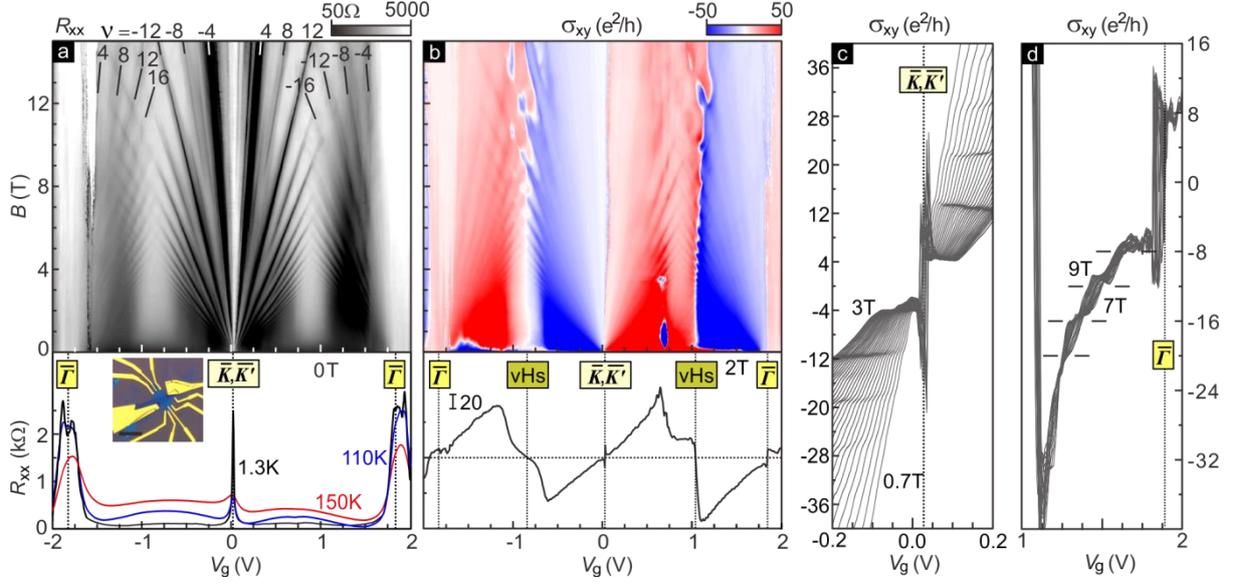

**Figure. 3**. Quantum Hall effect in twisted bilayer graphene. (a) Top: $R_{xx}$ as a function of gate voltage $V_g$ and magnetic field $B$. Quantum Hall states are highlighted by solid lines and marked with the associated integer filling factor. Bottom: Longitudinal resistance $R_{xx}$ as a function of gate voltage measured at three different temperatures in the absence of a magnetic field. The resistance peak near 0 V stems from the charge neutrality at the ($\bar{K}, \bar{K}'$) Dirac points, while the peaks near ±1.9 V can be assigned to the $\bar{\Gamma}$ charge neutrality points. The inset shows an optical image of the measured device. The scale bar corresponds to 10 μm. (b) Top: Color plot of the Hall conductivity $\sigma_{xy}$ in the ($V_g$, $B$)-plane. Bottom: Single line trace of $\sigma_{xy}$ at 2 T. Each conversion of the charge carriers from electrons to holes or vice versa is marked by a vertical dotted line together with the associated location in the band structure or Brillouin zone. (c and d) Hall conductivity versus backgate voltage $V_g$ for different magnetic fields ranging from 0.7 to 3 T (c), and 7 to 9 T (d) in 0.1 T steps.

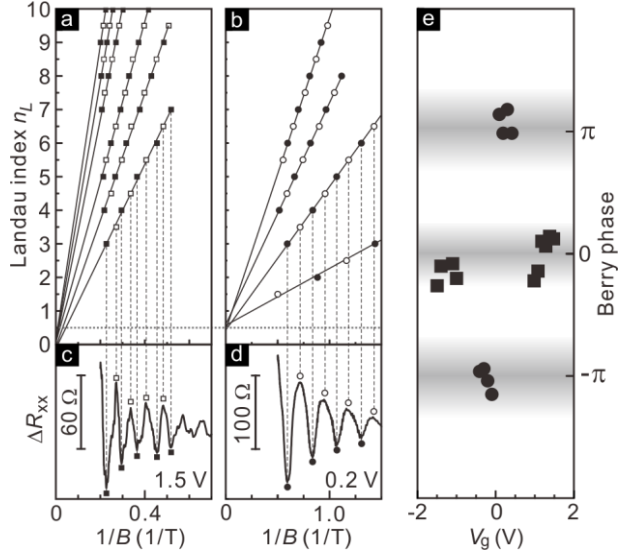

**Figure 4.** Berry phase determination from Shubnikov-de Haas extrema. (a) Fan diagram plotting the 1/B-field location of Shubnikov-de Haas extrema (filled squares for minima and empty squares for maxima) versus the Landau index $n_L$ for data recorded at gate voltages between 1.0 V to 1.5 V in 0.1 V steps. (b) The same as a but for $V_g$ between 0.1 V to 0.4 V (filled circles for minima, empty circles for maxima). (c) Examples of the Shubnikov-de Haas oscillations for $V_g = 1.5$ V and (d) $V_g = 0.2$ V at 1.3 K. (e) The intercept of linear fits to the data points in panel a and b with the Landau index axis multiplied by $2\pi$ yields the Berry phase. Its variation as a function of $V_g$ is plotted in this panel.

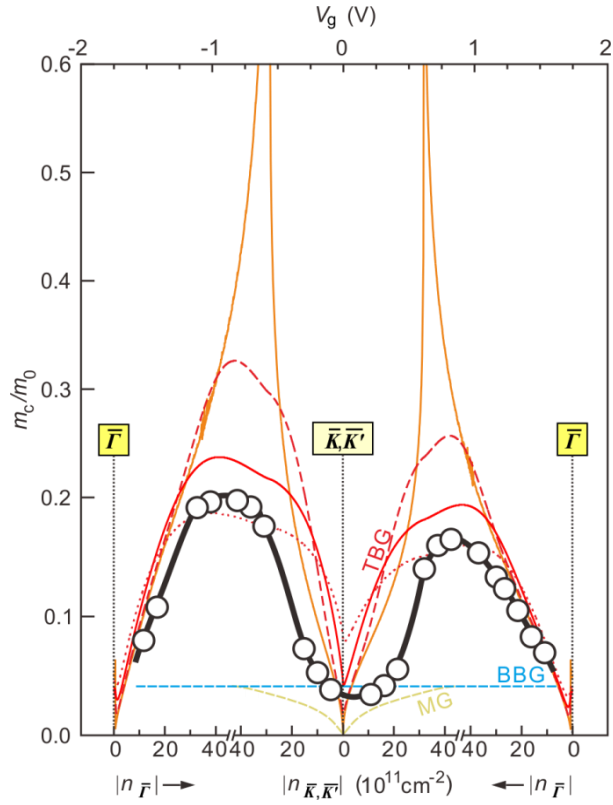

**Figure 5.** Evolution of the cyclotron mass with density or gate voltage. Black circles are the cyclotron mass extracted from the measured magnetotransport data. The thick solid black line serves as a guide to the eye. Other lines in this graph depict the theoretical dependencies of the cyclotron mass for monolayer graphene (MG, dashed dark yellow), Bernal stacked bilayer graphene (BBG, dashed blue line) and twisted bilayer graphene (TBG) having a twist angle of $\theta = 2°$ without any broadening (orange solid line) and with Gaussian broadening of varying width: a FWHM of 5 meV (dashed red line), 10 meV (solid red line), and 20 meV (dotted red line).

*Supporting Information for*

# Charge inversion and topological phase transition at a twist angle induced van Hove singularity of bilayer graphene

Youngwook Kim,[†] Patrick Herlinger,[†] Pilkyung Moon,[‡,§] Mikito Koshino,[∥] Takashi Taniguchi,[⊥] Kenji Watanabe,[⊥] and Jurgen. H. Smet[*,†]

[†]*Max-Planck-Institut für Festköperforschung, 70569 Stuttgart, Germany.*

[‡]*New York University, Shanghai 200120, China*

[§]*NYU-ECNU Institute of physics at NYU Shanghai, Shanghai 200062, China*

[∥]*Department of Physics, Tohoku University, Sendai 980-8578, Japan*

[⊥]*National Institute for Materials Science, 1-1 Namiki, Tsukuba, 305-0044, Japan*

**S1. Device images and hBN layer thickness determination**

Fig. S1 provides a schematic of the sequence of layers in our device. Also shown are optical microscope images of the device and contact geometry. Atomic force microscopy has been used to estimate the thickness of the hBN layers separating the twisted bilayer graphene from the graphitic gate layers. These measurements were performed on remaining fragments of the original, larger hBN layers. The fragments were separated from the main device during an etching step in the course of the device fabrication. The locations of the hBN fragments have been marked on the optical microscope image. The extracted thickness of these hBN layers is 6 nm (above the twisted bilayer) and 9 nm (below the twisted bilayer).

Since the top and bottom graphitic gate layers were accidently short circuited during fabrication, an electric field will develop across the twisted graphene layer when a gate voltage is applied due to the asymmetry in the thickness of the hBN dielectrics. Strictly speaking a self-consistent calculation is required to estimate this electric field as it also creates a charge imbalance between the two graphene layers. Since the electric field is rather small a crude approximation, which represents an upper limit for the field, can be obtained when we assume that the densities in both layers are identical. Using 5 and 2 as the static relative permittivity of hBN and graphene respectively, the potential difference between the two graphene layers amounts to U = 11.7 meV for an applied gate voltage of 0.5 V. In an AB stacked graphene bilayer the neglected charge imbalance reduces U by 50% [33]. If we assume that this order of reduction is similar in twisted bilayer graphene, then U = 6 meV. This small value of the potential difference only has minor influence on the calculated band structure, because interlayer interactions suppress shifts of the Dirac points[27].

## S2. Temperature dependence of the longitudinal resistance $R_{xx}$ at $B$ = 0 T

The gate voltage dependence of the longitudinal resistance in the absence of a magnetic field is shown in Fig. S2**a** for a set of 10 different temperatures. The temperature dependence of the resistance peaks near ± 1.9 V gate voltage and at the $(\bar{K}, \bar{K}')$ charge neutrality point (CNP) near zero gate voltage is presented in panel **b** on an inverse temperature axis. Note that for clarity the conductance $G_{xx} = 1/R_{xx}$ is plotted instead. While the conductance at the $(\bar{K}, \bar{K}')$ CNP decreases quickly when increasing the temperature, the other resistance maxima remain almost constant up to $T$ = 125 K ($1/T$ = 0.008 K$^{-1}$). They are therefore interpreted as a manifestation of the mini-gap between the lowest and the second conduction or valence bands. Their locations on the voltage axis are hereafter referred to as the $\bar{\Gamma}$ CNPs. We note however that these are indirect mini-gaps in momentum space only. The absolute minimum of the second bands are located along the $\bar{\Gamma}$ - $\bar{M}$ bar line. This can be seen from the contour plots of the second conduction bands in Fig. S3**a** as well as enlargements in Fig. S3**b** of the two-dimensional representation of the band dispersion close to the $\bar{\Gamma}$ –point of the lowest and second bands in Fig. 1. These bands exhibit the same mirror symmetry as the lowest conduction bands. Their absolute minimum overlaps slightly (2.9 meV) with the top of the lowest conduction bands at the $\bar{\Gamma}$ –point.

The complex shape of the bottom of the second bands presumably accounts for the broadness and double hump structure of these $\bar{\Gamma}$ CNP resistance features. In the inset to Fig. S2**b** we plot the temperature dependence of each resistance peak near the $\bar{\Gamma}$ CNPs on a logarithmic temperature scale. The largest peaks furthest away from the $(\bar{K},\bar{K}')$ CNP have been marked by a red empty square and a blue empty triangle in Fig. S2**a**, the other peaks by the corresponding filled symbols. All peaks show a similar temperature dependence that takes off only above 100 K. Two different channels may account and contribute to this temperature dependence. At low temperature, population of the second bands is only possible by tunneling into their absolute energy minimum. However the momentum mismatch effectively prevents or suppresses tunneling. At elevated temperature the phonon sphere expands. Acoustic phonon assisted tunneling enables these indirect transitions and the conductivity rises again[22,34]. Alternatively, direct transitions between the lowest and second band near the $\bar{\Gamma}$–point require bridging an energy gap of approximately 10 meV. An attempted fit of the temperature dependence of the resistance peaks to the expression for thermally activated behavior, despite the very limited temperature range available for such an analysis, yields activation energies between 9-12 meV. This activation energy range is comparable to the gap for direct transitions. We conjecture that both mechanisms are relevant in view of the double hump shape of the resistance features near the $\overline{\Gamma}$ CNPs.

**S3. Evidence for charge localization physics near the $\overline{\Gamma}$ –CNP**

Figure S4 shows the derivative of the longitudinal resistance with respect to the gate voltage in the magnetic field versus gate voltage plane. The resistance data themselves have been presented in Fig. 3**a**. The derivative reveals more clearly the Landau fans. Here we wish to highlight the vertical lines near the $\overline{\Gamma}$ –CNPs. They persist all the way down to zero magnetic field. Such bundles of parallel lines have been reported previously and were attributed to non-linear screening during charge localization when the electron system becomes incompressible[26,35]. A prerequisite for the observation of this physics is a gap in the energy spectrum. Similar to the observed thermally activated behavior of the resistance features discussed in S2, the appearance of these signatures for charge localization near the $\overline{\Gamma}$–CNPs also suggest the existence of a gap at B = 0 T.

**S4. Determination of the superlattice unit cell and twist angle**

In previous studies of van der Waals heterostructures[9-11,23,36,37] atomic force microscopy, scanning force microscopy or also Raman spectroscopy were exploited to assess the Moiré superlattice periodicity. In contrast to these previous reports, our device is encapsulated not only by h-BN but also graphite. Hence, all these methods cannot be conducted to measure the superlattice periodicity. The observation of charge neutrality at the $\overline{\Gamma}$ point in the longitudinal and Hall resistance offers however an alternative approach to estimate the superlattice periodicity. If we denote the carrier density for a completely filled single Bloch conduction band as $N_0$, the carrier density is equal to 1/$A$, where $A = \frac{\sqrt{3}}{2}a^2$ (Eq. 1) and $a$ is the superlattice unit cell length. Since there are two conductions bands and the charge carriers also possess a spin degree of freedom, complete filling up to the $\bar{\Gamma}$ CNP requires a total density of $4N_0$. The conversion factor between charge carrier density and gate voltage can be extracted from the slope of QH features and Shubnikov-de Haas oscillations. The relation between the gate voltage and carrier density is $N \approx \alpha \times V_g$, with $\alpha = 45\times10^{11}$ cm$^{-2}$/V. To move the chemical potential from the $(\bar{K}, \bar{K}')$ CNP to the $\bar{\Gamma}$–point we apply a gate voltage of 1.9 V and hence we conclude that the superlattice periodicity $a$ is approximately 7 nm. By means of the formula $a = \frac{a'}{2\sin(\theta/2)}$, where $a'$ is graphene lattice vector, we calculate a twist angle of 2°.

A second, alternative approach to estimate the size of the superlattice unit cell is based on the crossing of QH states of identical filling factor but originating from the different CNPs at $(\bar{K}, \bar{K}')$ and $\bar{\Gamma}$. With the help of the Wannier diagram, the Landau fan of QH states emanating from the $(\bar{K}, \bar{K}')$ and $\bar{\Gamma}$ CNPs can be understood from the Diophantine relation: $\left(\frac{N}{N_0}\right) = \nu\frac{\phi}{\phi_0} + s$ (Eq. 2). Here, $\nu$ is the filling factor, $\frac{\phi}{\phi_0}$ represents the magnetic field in dimensionless units by dividing the total flux $\phi = BA$ per unit cell with a magnetic flux quantum $\phi_0 = h/e$ and $s$ is the Bloch band filling index with $s = 0$ referring to states from the $(\bar{K}, \bar{K}')$ pocket and s= ± 4 for states of the $\bar{\Gamma}$ pockets. At the crossing of identical QH states from different CNPs $\frac{\phi}{\phi_0} = \frac{1}{q}$ with q an integer. For example, the $\nu_{(\bar{K},\bar{K}')}$ = -20 QH state for the $(\bar{K}, \bar{K}')$ CNP and $\nu_{\bar{\Gamma}}$ = 20 QH state for the secondary CNP at -1.9 V cross at B = 9.689 T as illustrated in Fig. S4. The use of Eq. 2 for both QH states yields $\left(\frac{N}{N_0}\right) = -20\frac{1}{q} = 20\frac{1}{q} - 4$ and hence $\frac{1}{q}$ is $\frac{1}{10} = \frac{BAe}{h}$. When substituting $A$ into Eq. 1 we obtain the superlattice unit cell length a = 7.021 nm. The corresponding twist angle equals 2.0°.

**S5. Carrier density analysis by Shubnikov-de Haas oscillations**

Figure S5**a** shows the Shubnikov-de Haas oscillations on an inverse magnetic field scale for a set of gate voltages for which the chemical potential is located either below or above the van Hove singularity. The periodicity of the oscillations can be deduced from a Fourier transform analysis as presented in panel **b**. The fundamental period and its second harmonic are marked by rectangular and star-shaped symbols, respectively. A special case occurs at 0.5 V. The Fourier transform of the oscillations reveals two peaks of comparable size. They are highlighted by a triangle and a rectangle. Near this gate voltage the layer degeneracy gets lifted by the transverse electric field generated due to the application of the gate voltage. This can also be observed in the two-dimensional rendition of the longitudinal resistance by the appearance of additional spokes in the Landau fan emanating from the $(\overline{K}, \overline{K}')$ CNP in Fig. 3a. The peak marked by the triangle is expected for an 8-fold degenerate 2D system, while the reduction of the degeneracy leads to a two times larger peak value marked by the rectangle. At the next gate voltage of 0.6 V only the peak corresponding to four fold degeneracy remains.

The Fourier analysis and extracted periodicities allow a determination of the charge carrier density as a function of gate voltage according to the expression $\Delta \frac{1}{B} = \frac{g}{nh} e$. Here, g is the degeneracy factor, *e* is elementary charge and *h* Planck's constant. The degeneracy factor depends on the gate voltage. Near the primary charge neutrality point, it equals 8 due to the layer, spin and valley degrees of freedom. The valley degeneracy accounts for the two mirror symmetric conduction bands originating from the $(K_t, K_b)$ and $(K'_t, K'_b)$ valley pairs of the two monolayer constituents, whereas the layer degeneracy reflects the co-existence of two iso-energetic closed orbits from the weakly coupled $\overline{K}$ and $\overline{K}'$ Dirac cones. Near 0.5 V the layer degeneracy is lifted due to the transverse electric field and the degeneracy factor reduces to 4. Beyond the van Hove singularity, the layer degeneracy is also non-existent due the hybridization of the Dirac cone states which results only in a single iso-energetic contour around the $\overline{\mathit{\Gamma}}$-point. The result of the density analysis is summarized in panel **c**. The charge carrier density drops to zero as we approach zero gate voltage and ± 1.9 V, corresponding to the primary CNP at the $(\overline{K}, \overline{K}')$ of the BZ and the secondary CNP at $\overline{\mathit{\Gamma}}$. The charge carrier density reaches its maximum near ± 0.9V. This suggests that the predominant charge carrier type is transformed from electrons to holes and vice versa at these gate voltages of the van Hove singularity. This is consistent with the Hall measurements presented in Fig. 3**b** which demonstrate that the rotation direction of the charge carriers is reversed at these voltages.

**S6. Cyclotron mass extraction from Shubnikov-de Haas oscillations**

As discussed in the main text, the cyclotron mass can be obtained by analyzing the temperature dependence of the extrema of Shubnikov-de Haas oscillations. Figure S6**a** presents an example at a gate voltage $V_g$ = 0.6 V. Plotted is the deviation of the longitudinal resistance from its value at zero magnetic field: $\Delta R_{xx} = R_{xx}(B) - R_{xx}(0)$. The temperature dependence of one resistance maximum (marked by the square symbol in panel **a**) further normalized to its value at 1.3 K is plotted in panel **b**. The data points can be fitted nicely with the Lifshitz-Kosevich formula, $\Delta R_{xx} \propto T/\sinh(2\pi^2 T m_c/\hbar eB)$, using the cyclotron mass, $m_c$, as the fitting parameter. The fit is included as the black solid line connecting the data points.

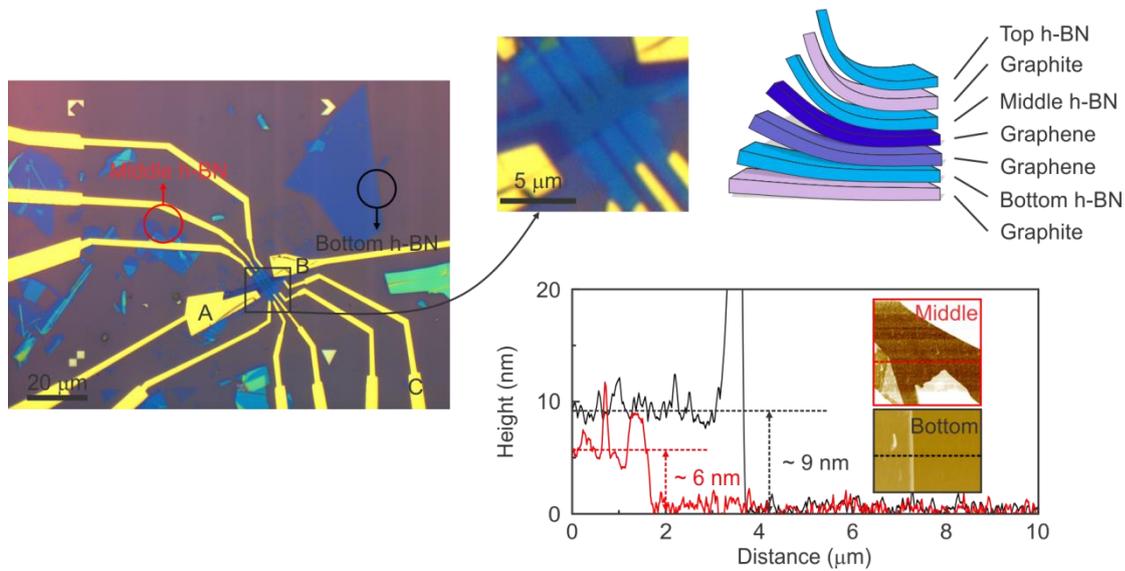

**Figure. S1.** Left: Optical micrograph of the measured device. The scale bar corresponds to 20 μm. The red (black) circle indicates the fragment of hBN flake immediately above (below) the twisted graphene bilayer. Two large gold electrodes (A,B) contact the graphitic back gate, while the graphitic top gate is connected via electrode C. The other electrodes connect to the twisted bilayer. Top right: Schematic illustration of the van der Waals heterostructure composed of seven layers. Bottom right: Thickness of the hBN flakes immediately below and above the twisted bilayer determined by AFM line scans. The red (black) line represents the AFM scan on a fragment of the middle hBN layer immediately above (below) the twisted bilayer graphene. The insets are color renditions of the topography in the regions that have been marked in the optical image. The dotted lines mark the positions where the line traces were recorded.

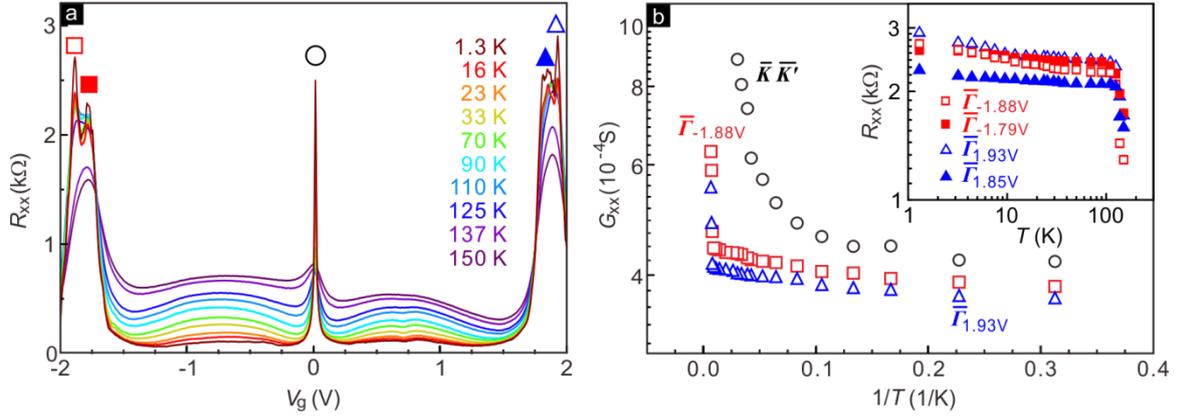

**Figure. S2 a,** Longitudinal resistance $R_{xx}$ as a function of gate voltage at 10 different temperatures. **b,** Temperature dependence at the $R_{xx}$ maxima for $V_g = 0.015$ V ($\bar{K}, \bar{K}'$ CNP) as well as $V_g = -1.88$ V and $+1.93$V ($\bar{\Gamma}$ CNP). Note that the conductance is plotted for clarity. The inset displays the temperature dependence of all $R_{xx}$ maxima in the double hump resistance features near the $\bar{\Gamma}$ CNPs on a logarithmic scale. They occur at $V_g = -1.88$ V, $-1.79$V, $1.85$V and $+1.93$V. The different symbols in panel **a** mark these peaks. The same symbols are used in panel **b**.

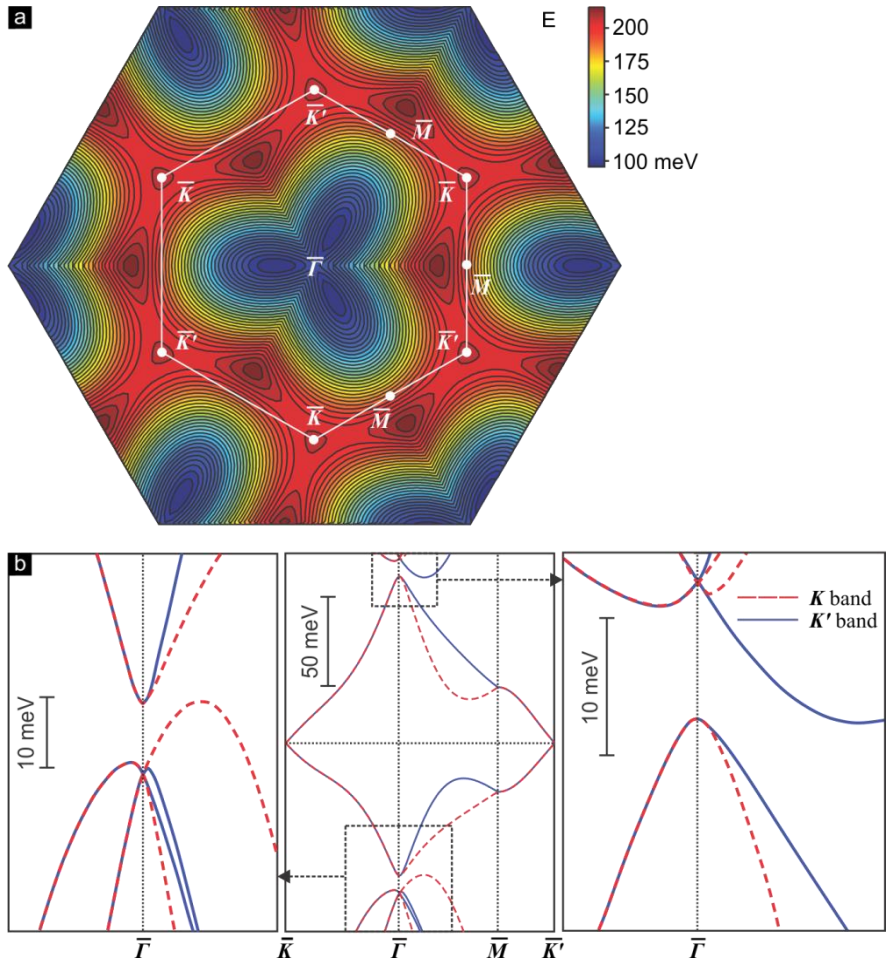

**Figure S3: a,** Contour plot for one of the second conduction bands. The contours are 4 meV apart. The other conduction band exhibits the same mirror symmetry as the lowest pair of conduction bands. The gaps between the lowest and second conduction bands are indirect. The absolute minimum occurs along the $\overline{\Gamma}$ - $\overline{M}$ direction. Direct transitions at the $\overline{\Gamma}$-point require an energy of approximately 10 meV. **b,** Zoom in of the conduction (right) and valence band (left) structure at the $\overline{\Gamma}$-point in order to highlight the dispersion of the bands that get filled after filling the lowest bands.

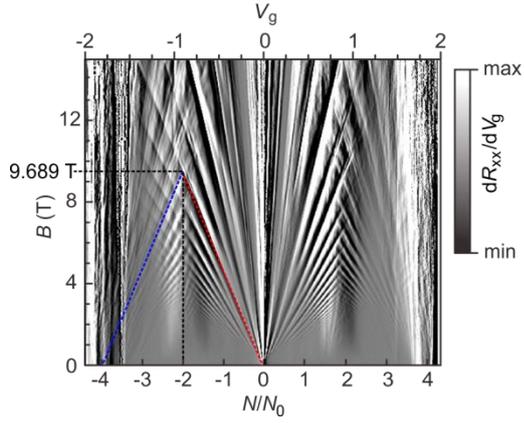

**Figure. S4** Derivative of the longitudinal magnetoresistance data with respect to the gate voltage in the top panel of Fig. 3b to bring out quantum Hall features more clearly. A bundle of vertical lines near +/-1.9 V running all the way down to zero field is reminiscent of charge localization physics when the system turns incompressible due to a gap in the energy spectrum. Red (blue) dot line indicates $v_{(\bar{K},\bar{K}')} = -20$ QH state for the $(\bar{K},\bar{K}')$ CNP ($v_{\bar{T}} = 20$ QH state for secondary CNP at -1.9 V, they cross at (9.689 T, $N/N_0 = -2$).

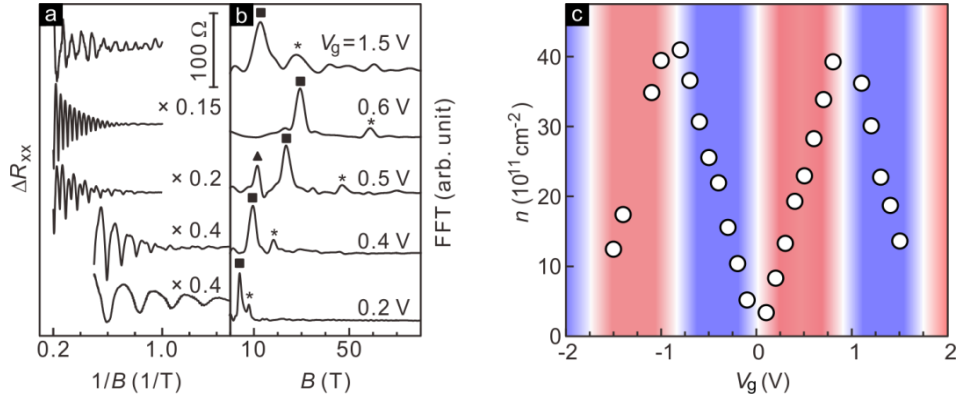

**Figure. S5 a,** Shubnikov-de Haas oscillations as a function of $1/B$ at different gate voltages. **b,** Fourier analysis of the Shubnikov-de Haas oscillations in order to extract the carrier density as a function of gate voltage. Square symbols mark the fundamental periodicity of the oscillations and stars the second harmonic. At a gate voltage of 0.5 V a transition occurs. Due to the transverse electric field, the layer degeneracy is lifted and the initial 8-fold degeneracy drops to 4-fold. As a result the second harmonics peak at 0.4 V has converted into the fundamental period at 0.5 V. A remnant of the original fundamental period is still visible but has vanished entirely at 0.6 V. **c,** Carrier density extracted from the Fourier analysis as a function of $V_g$.

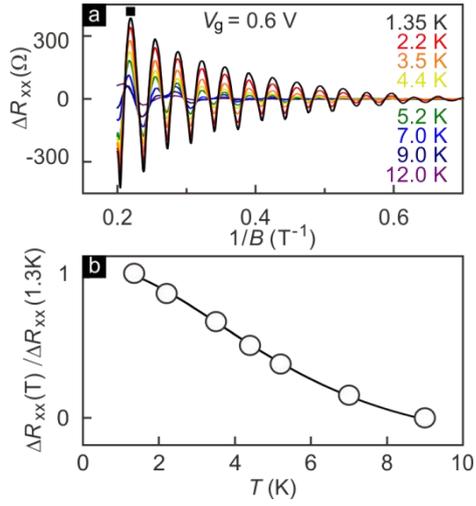

**Figure. S6 a,** $\Delta R_{xx} = R_{xx}(B) - R_{xx}(0)$ at $V_g = 0.6$ V at 8 different temperatures. **b,** $\Delta R_{xx}$ normalized by its value at 1.3K at the magnetic field of the largest maximum (marked by a square in panel **a)** as a function of temperature. The data is fitted by the Lifshitz-Kosevich formula. The fit is included as the solid line.